# A Bottom-Up Field-Theoretic Framework via Hierarchical Coarse-Graining: Generalized Mode Theory

Jaehyeok Jin,[1,2,#] Yining Han,[1,#] and Gregory A. Voth[1]*

[1]Department of Chemistry, Chicago Center for Theoretical Chemistry, Institute for Biophysical Dynamics, and James Franck Institute, The University of Chicago, Chicago, IL 60637, USA

[2]Department of Chemistry, Columbia University, New York, NY 10027, USA

[#]Authors contributed equally

* Corresponding author: gavoth@uchicago.edu

**Abstract**

Multiscale simulations facilitate the efficient exploration of large spatiotemporal scales in chemical and physical systems, yet particle-based simulations become prohibitively expensive at time and length scales beyond the molecular level. Field-theoretic simulations offer an attractive alternative, but most existing formulations rely on top-down approximations and are not systematically connected to atomistic interactions. Here, we present a hierarchical bottom-up framework for constructing auxiliary field representations of molecular liquids directly from microscopic models. We introduce a hierarchical coarse-graining framework that constructs field-theoretic models directly from atomistic liquids. The method first maps atomistic interactions to coarse-grained center-of-mass potentials and regularizes short-range divergences through a perturbative expansion in reciprocal space. Building on the auxiliary field formulation developed in polymer field-theoretic simulations, we then generalize the Hubbard-Stratonovich transformation to arbitrary pair potentials by separating positive and negative Fourier modes and introducing two auxiliary fields. The resulting generalized mode theory extends bottom-up field-theoretic modeling beyond positive-definite kernels and is compatible with existing field-theoretic sampling strategies. By combining formal derivations with numerical regularization and mode-truncation procedures, this work provides the theoretical foundation for scalable, bottom-up field-theoretic simulations of molecular systems.

## I. Introduction

In the study of molecular processes in chemical and physical systems, computer simulation has offered valuable insight along with efficiency and accuracy. Namely, atomistic particle-level simulations provide a molecular-level understanding of various chemical processes.[1-10] Extending the range of computer simulations to much longer time scales and larger length scales for complex molecular systems remains a challenging problem, yet "bottom-up" coarse-grained (CG) methodologies have been developed to address this issue by significantly reducing computational overheads.[5, 8, 10-18] Instead of simulating the fully atomistic representation of systems, CG methodologies aim to construct a reduced particle resolution defined by mapping many degrees of freedom to fewer ones, such as capturing collective motions,[19-22] charge,[23] symmetry,[24] etc. The effective CG interactions are then inferred from atomistic information, resulting in many-body CG potentials of mean force (PMFs).[21, 22] Examples of these methodologies include inverse Monte Carlo (IMC),[25, 26] iterative Boltzmann inversion (IBI),[27] relative entropy minimization (REM),[28-30] and multiscale coarse-graining (MS-CG) (aka "force-matching").[19-22, 31] In turn, bottom-up CG models provide a reduced representation of underlying atomistic models, enabling exploration of much larger spatiotemporal scales.

While bottom-up CG approaches can extend the range of computer simulations while retaining physical accuracy, they are still constructed at the particle level. A particle-level representation inherently and eventually faces computational limitations when extended to much larger and longer scales. For molecular systems with relatively high resolution, where CG entities correspond to groups of spatially coherent fine-grained (all-atom) entities, these limitations may be less pronounced due to theoretical advances in CG modeling, e.g., Ultra-Coarse-Graining (UCG),[32-37] as exemplified by recent studies in our group on virus capsid self-assembly.[38-43] and quantum mechanics/CG molecular mechanics (QM/MM).[44-46] However, this limitation becomes more pronounced at the mesoscopic and macroscopic levels beyond the micro-scales (i.e., μs and μm), where a specific CG entity may need to represent hundreds or even a much larger number of fine-grained degrees of freedom.[47-58] The larger spatial scale smears out local correlations, resulting in a more smoothly varying spatial representation of the molecular system. In such regimes, particle-based theories become less suitable, and field-theoretic approaches offer a more natural representation, consistent with how mesoscopic and macroscopic theories[59-61] such as fluid mechanics and continuum mechanics theories are fundamentally formulated.[62]

In principle, a long-term goal of bottom-up coarse-graining is to construct a well-defined field-theoretic model directly from a microscopic reference, with the ability to numerically sample the underlying fields. At the mesoscopic level and beyond, field variables represent highly CG entities whose correspondence to atomistic degrees of freedom is often indirect or unclear. As a result, most existing field-theoretic models are formulated phenomenological or top-down approaches, designed to reproduce experimental observables or capture essential physical behaviors in a phenomenological manner.[63] However, transforming from atomistic systems into field-theoretic representations presents a significant practical challenge in selecting appropriate field variables that can still faithfully encode microscopic physics.[64-67] Among possible choices, density fields offer the most direct connection to atomistic descriptions, as they can be formally defined via microscopic density operators. Indeed, field-theoretic treatments of density fields have been rigorously developed in classical density functional theory (DFT)[68, 69] and fluctuating hydrodynamics frameworks.[70] However, these approaches often require additional numerical procedures to determine density functionals[71, 72] or simulate fluctuating hydrodynamics, adding layers of complexity. In this work, we are particularly motivated by auxiliary-field formulations of polymer field theories, developed by Fredrickson and coworkers, where it is convenient to recast the problem in terms of the Fourier components of density fields.[73-78] In this formalism, the effective pairwise Hamiltonian relevant to polymer chains can be expressed as a quadratic form in reciprocal space, allowing the use of a Hubbard-Stratonovich (HS) transformation[79, 80] to introduce an auxiliary field for sampling the corresponding partition function. This approach has proven highly effective for capturing mesoscopic thermodynamics in soft matter systems, including block copolymers,[74, 81, 82] polyelectrolytes,[83-85] and complex fluids with mesoscopic structure,[77, 86] while still preserving aspects of the underlying microscopic physics through the simplified polymer representation.

Currently, there are two main challenges in extending auxiliary-field polymer field-theoretic approaches to complex molecular systems beyond polymers. First, unlike in polymer systems, where interactions are typically soft and can be approximated by simple analytical forms (e.g., Gaussian potentials,[87-90] Yukawa potentials,[91] Morse potentials,[92] or screened Coulombic potentials[93]), molecular liquids and biomolecular systems often exhibit strong, short-range repulsions and highly non-Gaussian interaction profiles that lack closed-form expressions. While such soft interactions allow for analytical field-theoretic treatments and capture coarse-grained behavior in many soft materials,[76, 77, 94] they fall short of representing the rich and often multiscale nature of real molecular interactions.[95, 96] For example, in polymer field-theoretic simulations, the absence of underlying microscopic features has been shown to limit the ability to capture fine structural effects in copolymers. Recent progress by Sherck et al.[97] and others[98-100] has shown that microscopically informed field theories, derived using bottom-up coarse-graining via relative entropy minimization, can successfully bridge particle-level and field-level physics. These studies demonstrated the capability of such models to capture large-scale coacervation and self-assembly behavior within polymeric systems, and they offer an important foundation for extending field-theoretic methods beyond traditional top-down formulations. At the same time, these formulations typically employ Gaussian interaction forms, which are well suited to polymer models[101] but are not intended to represent the broader range of interaction features encountered in large-scale heterogeneity of liquids[102] or macromolecules[103-107] with strongly repulsive cores.[108] In these systems, the Fourier transform of the pair potential can exhibit negative amplitudes or even diverge at short distances, especially for standard interaction types such as Lennard-Jones or Coulombic potentials, making direct density field formulations mathematically ill-defined. This problem is further exacerbated by the numerous atomistic types and degrees of freedom.

We approach this challenge through a hierarchical coarse-graining framework.[109] Rather than directly transforming the atomistic Hamiltonian into a field representation, we first introduce a molecular CG model as an intermediate level. This intermediate model renormalize the sheer number of atomistic degrees of freedom while preserving key structural and dynamical features, thereby offering a more tractable basis for constructing field variables. From this renormalized CG description, we aim to derive field representations that mitigate the complexity of statistical inference required to parameterize the field-theoretic Hamiltonian. The proposed hierarchical approach would allow for a more general CG framework that can accommodate non-Gaussian, non-analytical, and even divergent interactions while preserving fidelity to the underlying microscopic physics.

In line with the established auxiliary-field formulations developed by Fredrickson and co-workers for polymer systems,[73-78] this paper extends the range of auxiliary-field methods to the general coarse-graining of molecular liquids. In particular, as the first step of this bottom-up program, we extend the HS transformation to classical molecular systems within a hierarchical coarse-graining framework and clarify how microscopic pair interactions are represented at the field level. To formalize this connection, we follow the auxiliary-field formulation of Baeurle and Parrinello,[87] which, unlike the conventional polymer field-theoretic framework of Fredrickson and co-workers, is constructed directly for classical fluids with arbitrary pair potentials rather than for Gaussian-chain Hamiltonians. Although Fredrickson's framework would ultimately lead to a consistent field representation (as evidenced by Refs. 97-100), the Baeurle-Parrinello formulation provides a more natural starting point for bottom-up coarse-graining, where one works with an explicit effective pair potential $U(R)$ and its Fourier transform $U(k)$. Building on this formulation, we generalize the auxiliary-field representation to treat oscillatory and negative Fourier modes of the interaction kernel, and we analyze how this extension affects the resulting field theory. Starting from complex atomistic interactions involving many microscopic degrees of freedom, we first integrate out atomic coordinates to obtain effective center-of-mass pair potentials that retain essential molecular structure while greatly reducing the dimensionality. The extended auxiliary-field theory then provides a consistent mesoscopic field description of these coarse-grained interactions, thereby broadening the range of molecular systems accessible to bottom-up field-theoretic simulations.

## II. Generalized Mode Theory: Canonical Ensembles

### A. Auxiliary Field Formulations

To develop a hierarchical coarse-graining framework capable of handling general CG interactions, we begin by formalizing our extension to the auxiliary field approach commonly employed in self-consistent field theory (SCFT) and related methodologies. Specifically, we consider how a molecularly coarse-grained description, already reduced from atomistic resolution, can be further mapped onto a field-theoretic level. To establish this connection, we first review the auxiliary field formalism used in SCFT and then extend it to arbitrary CG interactions. This extension forms the foundational step for the hierarchical coarse-graining approach presented later in Sec. V.

To establish the framework, we focus initially on homogeneous, single-component molecular systems composed of $N$ particles with configurations $\mathbf{R}^N$. While the current focus is on single-component systems, the theory is not limited to this scope and will be generalized to multi-component systems in the companion paper, thereby extending the applicability of the proposed field-theoretic transformation. At the CG level, we define a well-established microscopic density field operator and its Fourier transform using delta functions as follows:

$$\rho(\mathbf{R}) = \sum_{I=1}^{N} \delta(\mathbf{R} - \mathbf{R}_I), \tag{1a}$$

$$\rho(\mathbf{k}) = \int d\mathbf{R}\, \rho(\mathbf{R}) \exp(-i\mathbf{k} \cdot \mathbf{R}) = \sum_{I=1}^{N} \exp(-i\mathbf{k} \cdot \mathbf{R}_I), \tag{1b}$$

in which the $\mathbf{k}$ vector denotes the reciprocal lattice vector that conforms to the periodic boundary conditions of the system of interest. Each component $k_\alpha$ (with $\alpha = x, y, z$) corresponds to discrete values defined by the periodic boundary conditions of a system with dimensions $L_x \times L_y \times L_z$, conforming to $k_\alpha = 2\pi n_\alpha / L_\alpha$, where $n_\alpha$ is an integer. To express the CG interaction energy $U(\mathbf{R}^N)$, we first assume that the pairwise potential $U(R)$ is well-defined at zero separation. While this condition does not generally hold for molecular systems, we will be revisited in Sec. V. Under this assumption, the total interaction energy can be written as:

$$U(\mathbf{R}^N) = \frac{1}{2}\sum_{I \neq J} U(\mathbf{R}_I - \mathbf{R}_J) = \frac{1}{2}\sum_{I,J} U(\mathbf{R}_I - \mathbf{R}_J) - \frac{1}{2}NU(0). \tag{2}$$

Here, the self-interaction term $\frac{1}{2}NU(0)$ is subtracted for notational consistency, ensuring that only the physical interactions ($I \neq J$) are retained. Alternatively, $U(\mathbf{R}^N)$ can be expressed in reciprocal space using the definition of the local density operator:

$$\begin{aligned} U(\mathbf{R}^N) &= \frac{1}{2}\int d\mathbf{R}d\mathbf{R}'\rho(\mathbf{R})U(\mathbf{R}-\mathbf{R}')\rho(\mathbf{R}') - \frac{1}{2}NU(0) \\ &= \frac{1}{2}\int d\mathbf{R}d\mathbf{R}'\rho(\mathbf{R})e^{i\mathbf{k}\cdot\mathbf{R}}\sum_{\mathbf{k}} U(\mathbf{k})\rho(\mathbf{R}')e^{-i\mathbf{k}\cdot\mathbf{R}'} - \frac{1}{2}NU(0) \\ &= \frac{1}{2}\sum_{\mathbf{k}} \rho^*(\mathbf{k})U(\mathbf{k})\rho(\mathbf{k}) - \frac{1}{2}NU(0), \end{aligned} \tag{3}$$

where $\Sigma_{\mathbf{k}}$ is shorthand for summation over all reciprocal-space vectors $\mathbf{k} = (k_x, k_y, k_z)$.

In this reciprocal-space representation, we note that the free energy function can be represented as a quadratic function in terms of the Fourier modes of the density field operator. This quadratic form of Hamiltonian imparts a Gaussian structure to the partition function $[\exp(-\beta U(\mathbf{R}^N))]$, making it amenable to the well-known HS transformation[79] to transform the molecular CG partition function to a field representation,

$$\exp\left(-\frac{1}{2}ax^2\right) = \frac{1}{\sqrt{2\pi a}}\int dy \exp\left(-\frac{y^2}{2a} - ixy\right). \tag{4}$$

Here, $a$ is a positive real constant. Notably, Eq. (4) effectively decouples pairwise intermolecular interactions from intramolecular ("bonded") interactions, as the HS transformation applies specifically to terms involving $\rho^*(\mathbf{k})U(\mathbf{k})\rho(\mathbf{k})$. In typical CG models, bonded terms are often modeled as harmonic interactions (bonds, angles, and dihedrals) in the direct coordinate integral. These terms straightforwardly yield Gaussian integrals in the partition function, which remain unaffected by the transformation. For completeness, we provide an illustrative derivation of how bonded interactions can be handled for multiple-site CG models of liquids in Appendix A. This derivation gives the bond constraint analogous to the Gaussian chain term in polymer field theory.[73, 110] While the main text primarily focuses on single-site CG models with no bonded terms for simplicity, our field-theoretic formulation of molecular CG partition functions is general and can accommodate multi-site CG representations.

The specific form of the HS transformation [Eq. (4)] introduces two significant challenges when generalizing this approach to broad range of molecular CG systems. First, the $a > 0$ condition restricts the transformation to CG interactions with strictly positive Fourier modes. Since most CG methods target interactions in real space, this constraint limits the applicability of the HS transformation to a narrow class of analytical potentials[111-113] with positive-definite Fourier modes. However, it is generally unlikely that Fourier components of atomistically-derived CG interactions will be entirely positive. Second, $\exp(-ixy)$ in the integrand of the transformation involves a complex exponential. When statistically evaluating such integrals using conventional numerical techniques, e.g., Monte Carlo sampling, it introduces a numerical phase problem (similar to sign problem[114-116] but more severe). Together, these issues have posed critical bottlenecks in the application of bottom-up coarse-graining frameworks to field-theoretic models. Here, we will first discuss the limited scenario where $\forall\mathbf{k},\ U(\mathbf{k}) > 0$ and then generalize it to the case where $U(\mathbf{k})$ can take arbitrary values and signs.

**C. Cases of Positive Fourier Modes**

Under the pairwise approximation, we begin by presenting the central result of Ref. 87, following their derivation as a formal starting point. While their treatment provides a useful foundation, we will introduce several clarifications and modifications that ultimately lead to a different form. Specifically, we begin with the partition function of a molecular system in the canonical ensemble under constant $NVT$ conditions, i.e., $\mathcal{Z}(N,V,T)$, given by

$$\begin{aligned}\mathcal{Z}(N,V,T) &= \frac{1}{\lambda_B^{3N} N!}\int d\mathbf{R}^N \exp\left(-\beta U(\mathbf{R}^N)\right) \\ &= \frac{1}{\lambda_B^{3N} N!}\int d\mathbf{R}^N \exp\left(-\frac{1}{2}\beta\sum_{\mathbf{k}}\rho^*(\mathbf{k})U(\mathbf{k})\rho(\mathbf{k}) + \frac{1}{2}\beta N U(0)\right),\end{aligned} \tag{5}$$

in which $\lambda_B$ represents the thermal de Broglie wavelength containing the momentum contribution in the form of Gaussian integrals. For the case when $\rho(\mathbf{k}) > 0$, we introduce an auxiliary field variable $\sigma$ to perform the HS transformation to $\mathcal{Z}(N,V,T)$. We emphasize that $\rho(\mathbf{k}) = \sum_{I=1}^{N} e^{-i\mathbf{k}\cdot\mathbf{R}_I}$ depends on all particle coordinates $\mathbf{R}^N = (\mathbf{R}_1, \cdots, \mathbf{R}_N)$. Throughout this paper, $\rho(\mathbf{k})$ should be interpreted as $\rho(\mathbf{k};\mathbf{R}^N)$, thus the exponential factor in Eq. (5) remains a function of $\mathbf{R}^N$ and cannot be taken outside the configurational integral $\int d\mathbf{R}^N$. For notational simplicity, however, we continue to denote it by $\rho(\mathbf{k})$.

Before applying the HS transformation, we first introduce a set of properties for real field variables in configuration space denoted as $A(\mathbf{R})$, satisfying $A(-\mathbf{k}) = A^*(\mathbf{k})$. These properties will be of use in the later derivations for specific field variables $\sigma(\mathbf{R})$, $U(\mathbf{R})$, and $\rho(\mathbf{R})$:

$$\sum_{\mathbf{k}\neq 0}\sigma(\mathbf{k})\rho^*(\mathbf{k}) = 2\sum_{\mathbf{k}>0}\left(\sigma^R(\mathbf{k})\rho^R(\mathbf{k}) + \sigma^I(\mathbf{k})\rho^I(\mathbf{k})\right), \tag{6a}$$

$$\sigma^*(\mathbf{k})\sigma(\mathbf{k}) = \sigma^*(-\mathbf{k})\sigma(-\mathbf{k}), \tag{6b}$$

$$\sum_{\mathbf{k}>0}\frac{1}{U(\mathbf{k})}\sigma^*(\mathbf{k})\sigma(\mathbf{k}) = \frac{1}{2}\sum_{\mathbf{k}\neq 0}\frac{1}{U(\mathbf{k})}\sigma^*(\mathbf{k})\sigma(\mathbf{k}). \tag{6c}$$

The HS transformation begins with the observation that the quadratic term in Eq. (3), except $\rho^*(\mathbf{k}=0)U(\mathbf{k}=0)\rho(\mathbf{k}=0)$, can be grouped as

$$
\begin{aligned}
\sum_{\mathbf{k}\neq 0} \rho^*(\mathbf{k})U(\mathbf{k})\rho(\mathbf{k}) &= \sum_{\mathbf{k}<0} \rho^*(\mathbf{k})U(\mathbf{k})\rho(\mathbf{k}) + \sum_{\mathbf{k}>0} \rho^*(\mathbf{k})U(\mathbf{k})\rho(\mathbf{k}) \\
&= \sum_{\mathbf{k}>0} |\rho(-\mathbf{k})|^2\, U(-\mathbf{k}) + \sum_{\mathbf{k}>0} |\rho(\mathbf{k})|^2\, U(\mathbf{k}) = \sum_{\mathbf{k}>0} |\rho(\mathbf{k})|^2\, [U(-\mathbf{k}) + U(\mathbf{k})] \\
&= 2\sum_{\mathbf{k}>0} |\rho(\mathbf{k})|^2 \mathrm{Re}\{U(\mathbf{k})\}\,,
\end{aligned}
\tag{7}
$$

due to $U(-\mathbf{k}) = U^*(\mathbf{k})$ for real-space pair potentials. Equation (7) suggests that only the real part of the Fourier transform of the pairwise interaction contributes to the total energy. Therefore, we define the real part $\phi(\mathbf{k}) \coloneqq \mathrm{Re}\{U(\mathbf{k})\}$, which is assumed to be positive in this case. Then, we observe that the exponential part contains two Gaussian contributions from the real part $\rho_R(\mathbf{k})$ and the imaginary part $\rho_I(\mathbf{k})$ of the Fourier modes of the density operator:

$$
\begin{aligned}
&\exp\left(-\frac{1}{2}\beta \sum_{\mathbf{k}\neq 0} \rho^*(\mathbf{k})U(\mathbf{k})\rho(\mathbf{k})\right) \\
&\qquad = \exp\left(-\beta \sum_{\mathbf{k}>0} |\rho(\mathbf{k})|^2\, \phi(\mathbf{k})\right) = \exp\left(-\beta \sum_{\mathbf{k}>0} (\rho_I(\mathbf{k})^2 + \rho_R(\mathbf{k})^2)\phi(\mathbf{k})\right).
\end{aligned}
\tag{8}
$$

Therefore, the HS transformation of the non-zero Fourier component involves auxiliary variables for $\rho_R(\mathbf{k})$, denoted as $\xi(\mathbf{k})$, and for $\rho_I(\mathbf{k})$, denoted as $\eta(\mathbf{k})$. Then, two HS transformations can be written as

$$
\begin{aligned}
&\exp\left(-\frac{1}{2}\beta \sum_{\mathbf{k}\neq 0} \rho^*(\mathbf{k})U(\mathbf{k})\rho(\mathbf{k})\right) \\
&\qquad = \prod_{\mathbf{k}>0} \frac{1}{4\pi\beta\phi(\mathbf{k})} \int \prod_{\mathbf{k}>0} d\xi(\mathbf{k})d\eta(\mathbf{k}) \exp\Bigg[-\sum_{\mathbf{k}>0} \frac{1}{4\beta\phi(\mathbf{k})} (\xi(\mathbf{k})^2 + \eta(\mathbf{k})^2) \\
&\qquad - i\sum_{\mathbf{k}>0} [\zeta(\mathbf{k})\rho^*(\mathbf{k})]\Bigg].
\end{aligned}
\tag{9}
$$

Instead of introducing two variables, for notational convenience, these two auxiliary fields can be compactly represented as a single complex (auxiliary) field $\zeta(\mathbf{k}) \coloneqq \xi(\mathbf{k}) + i\eta(\mathbf{k})$. Then, since the local density field $\rho(\mathbf{R})$ is a real field variable, there is an additional condition that $\xi$ and $\eta$ should satisfy, i.e., $\zeta(\mathbf{k}) = \zeta^*(-\mathbf{k})$. Now, we further simplify the functional form of Eq. (9). From Eq. (6), the last term in Eq. (9) becomes $i\sum_{\mathbf{k}>0} \zeta(\mathbf{k})\rho^*(\mathbf{k}) = i\sum_{\mathbf{k}>0}\big(\rho_R(\mathbf{k})\xi(\mathbf{k}) + \rho_I(\mathbf{k})\eta(\mathbf{k})\big)$, and $\prod_{\mathbf{k}>0} 1/4\pi\beta\phi(\mathbf{k}) = \prod_{\mathbf{k}\neq 0} \sqrt{1/4\pi\beta\phi(\mathbf{k})}$, since $\phi(\mathbf{k}) = \phi(-\mathbf{k})$ by definition. Therefore, we can absorb the factor of 2 into the integration variable to reduce the exponential integrand in the partition function into the following form.

$$
\exp\left(-\frac{1}{2}\beta\sum_{\mathbf{k}\neq 0}\rho^*(\mathbf{k})U(\mathbf{k})\rho(\mathbf{k})\right)
$$

$$
=\prod_{\mathbf{k}\neq 0}\left[2\sqrt{\frac{1}{\pi\beta\phi(\mathbf{k})}}\right]\int\prod_{\mathbf{k}>0}d\frac{\xi(\mathbf{k})}{2}d\frac{\eta(\mathbf{k})}{2}\exp\left[-\sum_{\mathbf{k}>0}\frac{1}{\beta\phi(\mathbf{k})}\left\{\left(\frac{\xi(\mathbf{k})}{2}\right)^2+\left(\frac{\eta(\mathbf{k})}{2}\right)^2\right\}\right.
$$

$$
\left.-2i\sum_{\mathbf{k}>0}\left[\rho_R(\mathbf{k})\left(\frac{\xi(\mathbf{k})}{2}\right)+\rho_I(\mathbf{k})\left(\frac{\eta(\mathbf{k})}{2}\right)\right]\right] \tag{10}
$$

Now, we define the final auxiliary field variable in a complex vector form $\sigma(\mathbf{k}) \coloneqq \sigma_R(\mathbf{k}) + i\sigma_I(\mathbf{k})$, with $\sigma_R(\mathbf{k}) = \xi(\mathbf{k})/2$ and $\sigma_I(\mathbf{k}) = \eta(\mathbf{k})/2$, which also satisfies $\sigma(\mathbf{k}) = \sigma^*(-\mathbf{k})$. The final result reads as

$$
\exp\left(-\frac{1}{2}\beta\sum_{\mathbf{k}\neq 0}\rho^*(\mathbf{k})U(\mathbf{k})\rho(\mathbf{k})\right)
$$

$$
=\prod_{\mathbf{k}\neq 0}\left[2\sqrt{\frac{1}{\pi\beta\phi(\mathbf{k})}}\right]\int\prod_{\mathbf{k}>0}d\sigma_R(\mathbf{k})d\sigma_I(\mathbf{k})\exp\left[-\frac{1}{2}\sum_{\mathbf{k}\neq 0}\frac{\sigma(\mathbf{k})\sigma^*(\mathbf{k})}{\beta\phi(\mathbf{k})}\right.
$$

$$
\left.-i\sum_{\mathbf{k}\neq 0}[\sigma(\mathbf{k})\rho^*(\mathbf{k})]\right]. \tag{11}
$$

Note that the last two summations in the exponential function are modified from $\Sigma_{\mathbf{k}>0}$ to $\Sigma_{\mathbf{k}\neq 0}/2$ in accordance with Eq. (6). In this final expression, our formulation differs from the equation originally presented in Ref. 87 by a factor of 2 for each reciprocal lattice vector. This underscores the importance of performing the HS transformation carefully over two distinct field variables. For canonical systems, the zero wave vector contribution does not require the HS transformation, as $\rho(\mathbf{k}=0) = N$ is a fixed quantity. Its contribution to the partition function is therefore simply $\exp\left[-\frac{\beta}{2}\phi(\mathbf{k}=0)N^2\right]$. Since $\phi(\mathbf{k}=0)$ is constant for a given interaction potential, it does not appear in the action functional (which only contains $\mathbf{k}\neq 0$ components in canonical cases).

Putting these elements together, the complete form of a canonical partition function [Eq. (5)] is expressed as an integral of two auxiliary fields as

$$
\mathcal{Z}(N,V,T) = \frac{1}{\lambda_B^{3N}N!}\int d\mathbf{R}^N\exp\left[\frac{1}{2}N\beta U(0)\right]\exp\left[-\frac{\beta}{2}\phi(\mathbf{k}=0)N^2\right]\prod_{\mathbf{k}\neq 0}\left(\frac{4}{\pi\beta\phi(\mathbf{k})}\right)^{\frac{1}{2}}
$$

$$
\times\int\prod_{\mathbf{k}>0}d\sigma_R(\mathbf{k})d\sigma_I(\mathbf{k})\exp\left[-\frac{1}{2}\sum_{\mathbf{k}}\frac{\sigma^*(\mathbf{k})\sigma(\mathbf{k})}{\beta\phi(\mathbf{k})}-i\sum_{\mathbf{k}}\sigma(\mathbf{k})\rho^*(\mathbf{k})\right]
$$

$$
=\frac{1}{\lambda_B^{3N}N!}\exp\left[\frac{1}{2}N\beta U(0)\right]\prod_{\mathbf{k}\neq 0}\left(\frac{4}{\pi\beta\phi(\mathbf{k})}\right)^{\frac{1}{2}}\int\prod_{\mathbf{k}>0}d\sigma(\mathbf{k})\exp\left[-\frac{1}{2}\sum_{\mathbf{k}}\frac{\sigma^*(\mathbf{k})\sigma(\mathbf{k})}{\beta\phi(\mathbf{k})}\right]
$$

$$\int d\mathbf{R}^N \exp\left[-i\sum_{\mathbf{k}} \sigma(\mathbf{k})\rho^*(\mathbf{k})\right],$$

(12)

where $\exp\left[\frac{1}{2}N\beta U(0)\right]$ arises from the self-interaction term in the CG Hamiltonian. In the first line of Eq. (12), the configuration integral $\int d\mathbf{R}^N$ remains present even after introducing the auxiliary fields $\sigma(\mathbf{k})$ via the HS transformation [Eq. (11)]. This is because the integration variables in the functional integral do not depend on $\mathbf{R}^N$, whereas the microscopic density modes $\rho(\mathbf{k})$ do. This $\mathbf{R}^N$-dependence can be made explicit and used to evaluate the configuration integral, ultimately yielding a purely functional integral over the auxiliary fields using the following property:

$$\int d\mathbf{R}^N \exp\left(-i\sum_{\mathbf{k}} \sigma(\mathbf{k})\rho^*(\mathbf{k})\right) = \int d\mathbf{R}^N \exp\left(-i\sum_{\mathbf{k}} \sigma(\mathbf{k})\sum_{I} \exp(i\mathbf{k}\cdot\mathbf{R}_I)\right)$$
$$= \int d\mathbf{R}^N \exp\left(-i\sum_{I} \sigma(\mathbf{R}_I)\right) = \left(\int d\mathbf{R} \exp\big(-i\sigma(\mathbf{R})\big)\right)^N.$$

(13)

Note that Eq. (13) contains some imaginary values in the exponential function, and in principle, this will give rise to the numerical phase problem in sampling this integral. We will now refer to this integral as a *phase integral* and later discuss how this phase problem is exacerbated in the case of grand canonical ensembles. From Eqs. (12) and (13), one can finally obtain an expression of the canonical partition function as a functional integral with respect to an auxiliary field

$$\mathcal{Z}(N,V,T) = \frac{1}{\lambda_B^{3N} N!}\int d\mu\big(\sigma(\mathbf{R})\big)\exp(-\mathcal{S}[\sigma(\mathbf{R})]),$$

(14)

in which the integral measure is defined as

$$d\mu[\sigma] = \exp\left(\frac{1}{2}N\beta\phi(0)\right)\exp\left[-\frac{\beta}{2}\phi(\mathbf{k}=0)N^2\right]\prod_{\mathbf{k}\neq 0}\left(\frac{4}{\pi\beta\phi(\mathbf{k})}\right)^{\frac{1}{2}}\prod_{\mathbf{k}>0} d\sigma_R(\mathbf{k})d\sigma_I(\mathbf{k}),$$

(15)

and the effective action functional $\mathcal{S}[\sigma(\mathbf{R})]$ takes the following form,

$$\mathcal{S}[\sigma(\mathbf{R})] = \frac{1}{2}\sum_{\mathbf{k}\neq 0}\frac{\sigma^*(\mathbf{k})\sigma(\mathbf{k})}{\beta\phi(\mathbf{k})} - N\ln\int d\mathbf{R}\, e^{-i\sigma(\mathbf{R})}.$$

(16)

In turn, for positive Fourier modes, Eqs. (14)-(16) suggest that the HS transformation can be readily applied to derive the field-theoretic description of the CG Hamiltonian. This approach is numerically advantageous, as the computational overhead depends only on the number of $k$-points, $n_k$, rather than the number of particles $N$. Therefore, consistent with the efficacy of auxiliary-field polymer field-theoretic methods, this method allows for efficiently describing the effective partition function at the field-theoretic level for large-scale molecular systems exhibiting non-trivial correlations where $N \gg n_k$.

**D. General Fourier Modes: Generalized Mode Theory**

The findings of the Subsection II B cannot be directly applied to the case of negative weights $a < 0$ in the $\exp\left(-\frac{1}{2}ax^2\right)$ form. Nevertheless, if $a < 0$, one can instead perform the HS transformation on $-a$, leading to the following relationship:

$$\exp\left(-\frac{1}{2}ax^2\right) = \frac{1}{\sqrt{2\pi|a|}}\int dy \exp\left(-\frac{y^2}{2|a|} + xy\right). \tag{17}$$

Equation (17) indicates that the negative Fourier modes of the pairwise interaction can also be transformed by the modified HS transformation. To generalize this approach for treating arbitrary Fourier modes that may include mixed positive and negative modes, our central claim is that one can decompose the real part of the Fourier modes of the pairwise interaction into two components $\phi(\mathbf{k}) = \nu(\mathbf{k}) + \omega(\mathbf{k})$, where $\nu(\mathbf{k})$ is always positive, while $\omega(\mathbf{k})$ remains negative for any reciprocal lattice vector $\mathbf{k}$. This decomposition is unique given the Fourier-transformed $\phi(\mathbf{k})$ modes. By performing this decomposition, the CG Hamiltonian in reciprocal space can be expressed as:

$$\begin{aligned} U(\mathbf{R}^N) &= \frac{1}{2}\sum_{\mathbf{k}} \rho^*(\mathbf{k})U(\mathbf{k})\rho(\mathbf{k}) - \frac{1}{2}NU(0) \\ &= \frac{1}{2}\sum_{\mathbf{k}} \rho^*(\mathbf{k})\big(\nu(\mathbf{k}) + \omega(\mathbf{k})\big)\,\rho(\mathbf{k}) - \frac{1}{2}NU(0) \end{aligned} \tag{18}$$

Our generalized mode theory can transform arbitrary CG interactions into a field-theoretic model by introducing an auxiliary field variable. In the case of the positive mode component, $\nu(\mathbf{k})$, the HS transformation yields two auxiliary field variables $\sigma_R(\mathbf{k})$ and $\sigma_I(\mathbf{k})$ for $\mathbf{k} \neq 0$, as discussed in Sec II B:

$$\begin{aligned} \exp\left[-\frac{1}{2}\beta\sum_{\mathbf{k}\neq 0} \rho^*(\mathbf{k})\nu(\mathbf{k})\rho(\mathbf{k})\right] &= \prod_{\mathbf{k}\neq 0}\left(\frac{4}{\pi\beta\nu(\mathbf{k})}\right)^{\frac{1}{2}} \int \prod_{\mathbf{k}>0} d\sigma_R(\mathbf{k})d\sigma_I(\mathbf{k}) \\ &\times \exp\left[-\frac{1}{2}\sum_{\mathbf{k}\neq 0}\frac{\sigma^*(\mathbf{k})\sigma(\mathbf{k})}{\beta\nu(\mathbf{k})} - i\sum_{\mathbf{k}\neq 0}\sigma(\mathbf{k})\rho^*(\mathbf{k})\right]. \end{aligned} \tag{19}$$

For the negative part, the modified HS transform [Eq. (17)] is applied to the second auxiliary field $\xi(\mathbf{k}) \coloneqq \xi_R(\mathbf{k}) + i\xi_I(\mathbf{k})$, giving

$$\begin{aligned} \exp\left[-\frac{1}{2}\beta\sum_{\mathbf{k}\neq 0} \rho^*(\mathbf{k})\omega(\mathbf{k})\rho(\mathbf{k})\right] &= \prod_{\mathbf{k}\neq 0}\left(\frac{4}{\pi\beta|\omega(\mathbf{k})|}\right)^{\frac{1}{2}} \int \prod_{\mathbf{k}>0} d\xi_R(\mathbf{k})d\xi_I(\mathbf{k}) \\ &\times \exp\left[-\frac{1}{2}\sum_{\mathbf{k}\neq 0}\frac{\xi^*(\mathbf{k})\xi(\mathbf{k})}{\beta|\omega(\mathbf{k})|} + \sum_{\mathbf{k}\neq 0}\xi(\mathbf{k})\rho^*(\mathbf{k})\right]. \end{aligned} \tag{20}$$

Similar to the positive mode case, the canonical partition function does not explicitly require the zero wave vector contribution of $\phi(\mathbf{k} = 0)$, but the underlying reasoning is slightly different. While the generalized Fourier mode decomposition implies that one should consider both $\nu(\mathbf{k})$ and $\xi(\mathbf{k})$, i.e., $\phi(\mathbf{k} = 0) = \nu(\mathbf{k} = 0) + \xi(\mathbf{k} = 0)$, at the zero wave vector, either $\nu(\mathbf{k} = 0)$ or $\xi(\mathbf{k}=0)$ vanishes to zero due to its sign, i.e., $\exp(0)=1$. Hence, it reduces to the case of the positive Fourier mode, and then the remaining factor becomes

$$\exp\left[-\frac{\beta}{2}\big(\nu(\mathbf{k}=0)+\xi(\mathbf{k}=0)\big)N^2\right]=\exp\left[-\frac{\beta}{2}\phi(\mathbf{k}=0)N^2\right]. \tag{21}$$

By combining Eqs. (19) and (20) with the zero wave vector contribution, we arrive at the bottom-up field-theoretic representation of the canonical partition function, expressed as:

$$\mathcal{Z}(N,V,T)=\frac{1}{\lambda_B^{3N}N!}\exp\left(\frac{1}{2}\beta NU(0)\right)\exp\left[-\frac{\beta}{2}\phi(\mathbf{k}=0)N^2\right]\prod_{\mathbf{k}\neq 0}\left[\left(\frac{4}{\pi\beta\nu(\mathbf{k})}\right)^{\frac{1}{2}}\cdot\left(\frac{4}{\pi\beta|\omega(\mathbf{k})|}\right)^{\frac{1}{2}}\right]$$
$$\times\int\mathcal{D}\sigma(\mathbf{R})\mathcal{D}\xi(\mathbf{R})\left\{\exp\left[-\frac{1}{2}\sum_{\mathbf{k}\neq 0}\left(\frac{\sigma^*(\mathbf{k})\sigma(\mathbf{k})}{\beta\nu(\mathbf{k})}+\frac{\xi^*(\mathbf{k})\xi(\mathbf{k})}{\beta|\omega(\mathbf{k})|}\right)\right]\times\left(\int d\mathbf{R}\exp\big(-i\sigma(\mathbf{R})+\xi(\mathbf{R})\big)\right)^N\right\}. \tag{22}$$

In Eq. (22), the integrands are abbreviated as $\mathcal{D}\sigma(\mathbf{R})\coloneqq\prod_{\mathbf{k}>0}d\sigma_R(\mathbf{k})d\sigma_I(\mathbf{k})$ and $\mathcal{D}\xi(\mathbf{R})\coloneqq\prod_{\mathbf{k}>0}d\xi_R(\mathbf{k})d\xi_I(\mathbf{k})$. To reveal the mathematical structure more clearly, Eq. (25) can be compactly expressed as

$$\mathcal{Z}(N,V,T)=\frac{1}{\lambda_B^{3N}N!}\int d\mu(\sigma,\xi)\exp(-S[\sigma,\xi]), \tag{23}$$

where the integration measure is defined as

$$d\mu(\sigma,\xi)\coloneqq\exp\left(\frac{1}{2}\beta NU(0)\right)\exp\left[-\frac{\beta}{2}\phi(\mathbf{k}=0)N^2\right]\prod_{\mathbf{k}\neq 0}\left[\left(\frac{4}{\pi\beta\nu(\mathbf{k})}\right)^{\frac{1}{2}}\cdot\left(\frac{4}{\pi\beta|\omega(\mathbf{k})|}\right)^{\frac{1}{2}}\right]\mathcal{D}\sigma(\mathbf{R})\mathcal{D}\xi(\mathbf{R}), \tag{24}$$

and the action functional can be written as

$$S[\sigma,\xi]=\frac{1}{2}\sum_{\mathbf{k}\neq 0}\left[\frac{1}{\beta\nu(\mathbf{k})}\sigma^*(\mathbf{k})\sigma(\mathbf{k})+\frac{1}{\beta|\omega(\mathbf{k})|}\xi^*(\mathbf{k})\xi(\mathbf{k})\right]-N\ln\int d\mathbf{R}\,e^{-i\sigma(\mathbf{R})+\xi(\mathbf{R})}. \tag{25}$$

It is worth noting that in the effective action functional defined in Eq. (25), the terms associated with the auxiliary field variable $\xi(\mathbf{R})$ in the phase integral are always real and correspond to the positive statistical weights. Therefore, they do not contribute to the numerical phase problem in Monte Carlo sampling. The primary numerical challenges arise from the positive Fourier modes.

## III. Generalized Mode Theory: Grand Canonical Ensembles

In field-theoretic simulations, grand canonical systems, where the number of particles changes, are more desirable. As demonstrated by Fredrickson and co-workers, grand-canonical field-theoretic simulations, including mean-field SCFT and beyond mean-field simulations, enable efficient sampling of composition fluctuations and access to a broader range of morphologies than is possible with conventional particle-based methods.[81, 82, 117] In particular, varying the chemical potential in grand-canonical SCFT allows one to trace equilibrium structures across different compositions without the severe sampling barriers associated with particle insertion at high densities in traditional grand-canonical Monte Carlo or molecular dynamics simulations.[9] Yet, the canonical description of the field-level partition function serves as a starting point for deriving the grand canonical-level description of classical field models.The general structure of extending the single-component case to a multi-component case remains invariant under the grand canonical ensembles (interaction profiles do not change). Therefore, in this section, we primarily discuss the single-component case for conciseness.

**A. Positive Fourier Modes**

Based on the correct canonical formulation [Eq. (25)], for positive-only modes, we followe Ref. 87 to derive correct grand canonical cases. Since the grand canonical partition function $\Theta(\mu, V, T)$ is derived based on the canonical partition function: $\Theta(\mu, V, T) = \sum_{N=0}^{\infty} e^{\beta\mu N}\, \mathcal{Z}(N, V, T)$, the mathematical structure of $\Theta(\mu, V, T)$ at the field-level involves extracting the Gaussian measure outside of $\Sigma_N$ and grouping $e^{\beta\mu N}$ with the self-energy term $\exp\left(\frac{1}{2}N\beta\phi(0)\right)$ and the phase integral, i.e., $\left(\int d\mathbf{R}\exp\left(-i\sigma(\mathbf{R})\right)\right)^N = \exp\left(N\ln\int d\mathbf{R}\, e^{-i\sigma(\mathbf{R})}\right)$. Importantly, for grand canonical systems, we can introduce the zero wave vector component of $\rho(\mathbf{k})$ (as the number of particles changes). In this case, the HS transformation of the zero mode becomes

$$\exp\left[-\frac{\beta}{2}\rho^*(\mathbf{k}=0)\phi(\mathbf{k}=0)\rho(\mathbf{k}=0)\right] = \frac{1}{\sqrt{2\pi\beta\phi(0)}}\int d\sigma(0)\cdots = \frac{1}{2\sqrt{2}}\sqrt{\frac{4}{\pi\beta\phi(0)}}\int d\sigma(0)\cdots, \tag{26}$$

This implies that the action functional and integration measure now spans all $\mathbf{k}$ vectors, including $\mathbf{k} = 0$.

In order to separate out the $N$-dependence, we define an $N$-independent integral measure [which is now possible because we decoupled $\sigma(0)$], $d\tilde{\mu}[\sigma]$, as $d\mu[\sigma]d\sigma(0) = \exp\left(\frac{1}{2}N\beta\phi(0)\right)d\tilde{\mu}[\sigma]$, giving

$$\Theta(\mu, V, T) = \int d\tilde{\mu}[\sigma(\mathbf{R})]\exp\left(-\frac{1}{2}\sum_{\mathbf{k}}\frac{\sigma^*(\mathbf{k})\sigma(\mathbf{k})}{\beta\phi(\mathbf{k})}\right)\sum_{N=0}^{\infty}\frac{1}{\lambda_B^{3N}N!}\exp\left[N\left(\beta\mu + \frac{1}{2}\beta\phi(0) + \ln\int d\mathbf{R}\, e^{-i\sigma(\mathbf{R})}\right)\right]. \tag{27}$$

We can further simplify Eq. (27) as follows. First, as $\mu, \lambda_B$, and $\phi(0)$ are constants or system-dependent coefficients, we can group them as a fugacity-like constant, $\chi$, defined as $\chi \coloneqq \exp\left(\beta\mu + \frac{1}{2}\beta\phi(0)\right)/\lambda_B^3$. Then, we then notice that a term inside $\Sigma_{N=0}^{\infty}$ is simply a Taylor expansion of $\exp(\chi\int d\mathbf{R}e^{-i\sigma(\mathbf{R})}) = \sum_N^{\infty}\frac{\chi^N}{N!}\left(\int d\mathbf{R}e^{-i\sigma(\mathbf{R})}\right)^N$.

Next is the integral measure $d\tilde{\mu}[\sigma]$. By incorporating the zero wave vector contribution, as noted in Eq. (26), we can simplify $\Theta(\mu, V, T)$ as

$$\Theta(\mu,V,T) \coloneqq \int \mathcal{D}\mu[\sigma(\mathbf{R})]\exp\left(-\frac{1}{2}\sum_{\mathbf{k}}\frac{\sigma^*(\mathbf{k})\sigma(\mathbf{k})}{\beta\phi(\mathbf{k})}\right)\exp\left(\chi\int d\mathbf{R}\, e^{-i\sigma(\mathbf{R})}\right)$$
$$\coloneqq \int \mathcal{D}\mu[\sigma(\mathbf{R})]\exp(-S_{\mathrm{GC}}[\sigma(\mathbf{R})]), \tag{28}$$

with $\mathcal{D}\mu[\sigma(\mathbf{R})] \coloneqq \frac{1}{2\sqrt{2}}\prod_{\mathbf{k}}\sqrt{\frac{4}{\pi\beta\phi(\mathbf{k})}}\,d\sigma(0)\prod_{\mathbf{k}>0} d\sigma_R(\mathbf{k})d\sigma_I(\mathbf{k})$. Note that the $1/2\sqrt{2}$ prefactor is due to $\mathbf{k}=0$ contribution. Alternatively, one can decouple the $\phi(\mathbf{k})$ dependence from the integral measure $\mathcal{D}\mu[\sigma(\mathbf{R})]$ by introducing a normalization constant for $\phi(\mathbf{k})$ defined as $\gamma_\phi = \frac{1}{2\sqrt{2}}\prod_{\mathbf{k}}\sqrt{\frac{4}{\pi\beta\phi(\mathbf{k})}}$, giving

$$\mathcal{D}\mu[\sigma(\mathbf{R})] = \gamma_\phi d\sigma(0)\prod_{\mathbf{k}>0} d\sigma_R(\mathbf{k})d\sigma_I(\mathbf{k}). \tag{29}$$

Note that the multiplication now involves all $\mathbf{k}$ vectors (whereas $\mathbf{k}\neq 0$ for the canonical case), and the effective action functional in grand canonical ensembles can now be expressed as

$$S_{\mathrm{GC}}[\sigma(R)] \coloneqq \frac{1}{2}\sum_{\mathbf{k}}\frac{\sigma^*(\mathbf{k})\sigma(\mathbf{k})}{\beta\phi(\mathbf{k})} - \chi\int d\mathbf{R}\, e^{-i\sigma(\mathbf{R})}. \tag{30}$$

The major difference between the grand canonical action and the canonical action is that the imaginary part is in an exponential form. This suggests that the phase problem for sampling this integral will be significantly more challenging compared to the canonical case, consistent with Ref. 87. Such numerical instability requires an effective numerical sampling method to further reduce statistical noise when sampling grand canonical field models.

### B. General Fourier Modes: Generalized Mode Theory

Now, we extend our findings from Sec. III A to the case of general Fourier modes by separating the Fourier-transformed interaction $\phi(\mathbf{k})$ into two parts: the positive-definite $\nu(\mathbf{k})$ and the negative-definite $\omega(\mathbf{k})$. This allows us to express the grand canonical partition function as:

$$\Theta(\mu,V,T) = \sum_{N=0}^{\infty} e^{\beta\mu N}\mathcal{Z}(N,V,T) = \int \mathcal{D}\mu[\sigma,\xi]\exp(-S_{\mathrm{GC}}[\sigma,\xi]). \tag{31}$$

The effective action functional with respect to the auxiliary fields $\sigma$ and $\xi$ is expressed as,

$$S_{\mathrm{GC}}[\sigma,\xi] \coloneqq \frac{1}{2}\sum_{\mathbf{k}}\left[\frac{1}{\beta\nu(\mathbf{k})}\sigma^*(\mathbf{k})\sigma(\mathbf{k}) + +\frac{1}{\beta|\omega(\mathbf{k})|}\xi^*(\mathbf{k})\xi(\mathbf{k})\right] - \chi\int d\mathbf{R}\, e^{-i\sigma(\mathbf{R})+\xi(\mathbf{R})}, \tag{32}$$

and the integral measure $\mathcal{D}\mu[\sigma,\xi]$ includes the zero wave vector contributions from both $\sigma(\mathbf{R})$ and $\xi(\mathbf{R})$ and is defined as

$$\mathcal{D}\mu[\sigma,\xi] \coloneqq \frac{1}{8}\prod_{\mathbf{k}}\left[\sqrt{\frac{4}{\pi\beta\nu(\mathbf{k})}}\cdot\sqrt{\frac{4}{\pi\beta|\omega(\mathbf{k})|}}\right]d\sigma(0)d\xi(0)\prod_{\mathbf{k}>0} d\sigma_R(\mathbf{k})d\sigma_I(\mathbf{k})d\xi_R(\mathbf{k})d\xi_I(\mathbf{k}). \tag{33}$$

Equations (31)-(33) are the main results of this subsection. Similar to Eqs. (23)-(25), we observe that $\omega(\mathbf{k})$ does not contribute to the phase problem, even though its magnitude increases exponentially. The primary challenge in implementing Eqs. (32) and (33) is to address the contribution from $\sigma(\mathbf{R})$ through $\nu(\mathbf{k})$.

## IV. Numerical Demonstration

### A. System Setup

To numerically assess the feasibility of the generalized mode theory, we consider a two-dimensional (2D) model system with radial interactions defined in Fourier space as

$$U(k) = A_+ \exp\left(-\frac{k^2}{2\sigma_+^2}\right) + A_- \exp\left(-\frac{(k-k_0)^2}{2\sigma_-^2}\right), \tag{34}$$

which naturally decomposes into a positive Gaussian mode $\nu(k) = A_+ \exp\left(-\frac{k^2}{2\sigma_+^2}\right)$ and a negative Gaussian ring mode $\omega(k) = A_- \exp\left(-\frac{(k-k_0)^2}{2\sigma_-^2}\right)$ [Fig. 1(a)]. Here, $k = |\mathbf{k}|$ is the radial wave number, and the parameters are $A_+ = 40$, $\sigma_+ = 1.0$, $A_- = -15$, $\sigma_- = 0.35$, and $k_0 = 2.0$ Å$^{-1}$. The choice of interaction form and parameters is motivated by its real-space counterpart $U(R)$, which exhibits the characteristic repulsive-attractive oscillatory profile commonly found in liquid interactions, as shown in Fig. 1(b). The real-space interaction $U(R)$ is obtained via the 2D inverse Hankel transform using the zeroth-order Bessel function of the first kind $J_0$, given by

$$U(R) = \int_0^\infty \frac{k\,dk}{2\pi} U(k) J_0(kR), \tag{35}$$

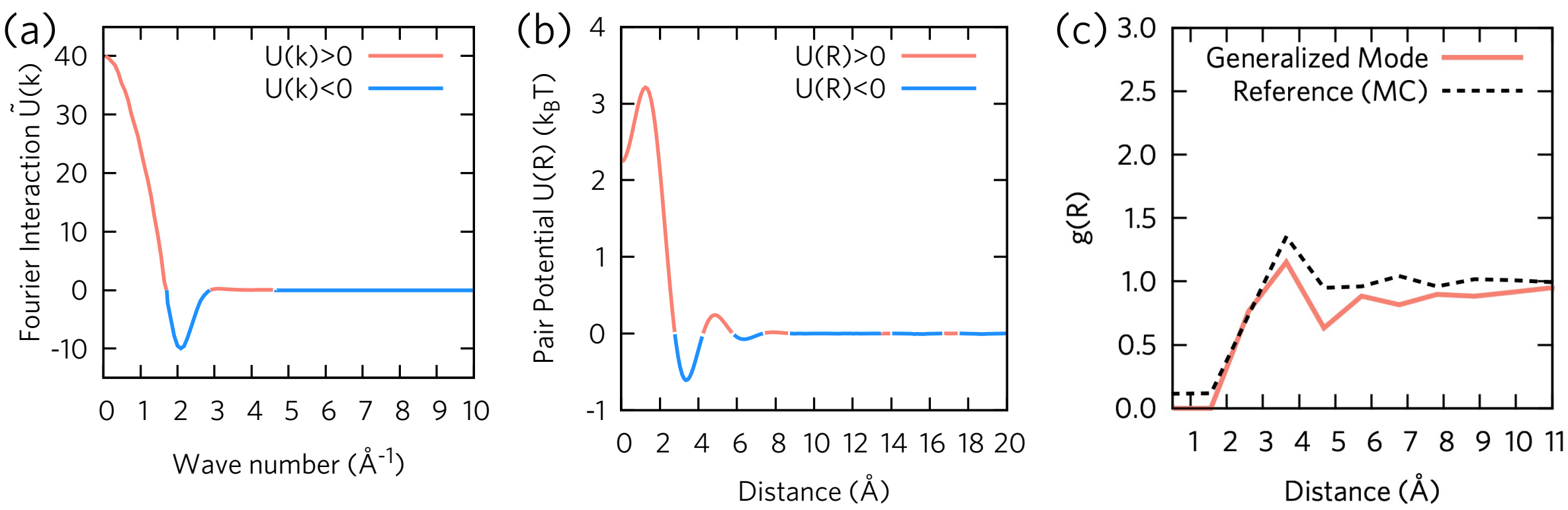

**Figure 1:** Numerical demonstration of the generalized mode theory using field-theoretic sampling. (a) Model interaction in reciprocal space $U(k)$ [Eq. (34)] with positive (red) and negative (blue) modes. (b) Real-space interaction $U(R)$ obtained from Eq. (35) with positive (red) and negative (blue) interactions. (c) 2D RDF computed from the generalized mode theory using panel (a) with complex Langevin sampling (red), compared against the real-space particle MC reference (black dashed).

### B. Mode-separated auxiliary fields evolution

To provide a minimal numerical demonstration of the generalized mode theory, we sample the auxiliary fields of the canonical partition function using complex Langevin sampling.[86, 118, 119] We consider a square box of length 50 Å with periodic boundary conditions, containing 64 particles. Complex Langevin sampling is performed on a uniform $48 \times 48$ spectral grid, with the wavevectors conforming to the periodic boundary condition:

$$\mathbf{k} = (k_x, k_y) = \frac{2\pi}{L}(n_x, n_y), \tag{36}$$

where $n_x, n_y \in [-24,24)$. We note that in the present work, we adopt a minimal spectral grid resolution sufficient to demonstrate the feasibility of the generalized mode formulation. Higher-resolution calculations, while desirable, are numerically demanding for oscillatory kernels and are left as an important direction for future work.

On this grid, the generalized mode theory splits $U(k)$ into positive and negative components:

$$v(k) = \max\{U(k), 0\} + \epsilon, \tag{37a}$$

$$|\omega(k)| = -\min\{U(k), 0\} + \epsilon, \tag{37b}$$

with $\epsilon = 10^{-12}$ to avoid singularities at $k = 0$. Two independent Hubbard-Stratonovich transformations are then introduced, associated with auxiliary fields $\sigma(\mathbf{k})$ (positive mode) and $\xi(\mathbf{k})$ (negative mode). In doing so, we write the generalized-mode Hamiltonian as

$$H[\sigma, \xi] = H_{\mathrm{G}}[\sigma, \xi] + H_{\mathrm{int}}[\sigma, \xi], \tag{38}$$

where $H_{\mathrm{G}}$ collects the quadratic Gaussian terms,

$$H_{\mathrm{G}} = \sum_{\mathbf{k}} \left[ \frac{|\sigma(\mathbf{k})|^2}{2\beta v(k)} + \frac{|\xi(\mathbf{k})|^2}{2\beta |\omega(k)|} \right], \tag{39}$$

and $H_{\mathrm{int}}$ denotes the non-Gaussian term arising from the single-particle partition function involving phase integral. The resulting generalized-mode Hamiltonian leads to two separate complex Langevin equations:

$$\frac{\partial \sigma(\mathbf{k}, \theta)}{\partial \theta} = -a_\sigma(\mathbf{k}) \sigma(\mathbf{k}, \theta) - \frac{\delta H_{\mathrm{int}}}{\delta \sigma^*(\mathbf{k}, \theta)} + \eta_\sigma(\mathbf{k}, \theta) \tag{40a}$$

$$\frac{\partial \xi(\mathbf{k}, \theta)}{\partial \theta} = -a_\xi(\mathbf{k}) \xi(\mathbf{k}, \theta) - \frac{\delta H_{\mathrm{int}}}{\delta \xi^*(\mathbf{k}, \theta)} + \eta_\xi(\mathbf{k}, \theta) \tag{40b}$$

with $a_\sigma(\mathbf{k}) = [\beta v(k)]^{-1}$, $a_\xi(\mathbf{k}) = [\beta |\omega(k)|]^{-1}$, $\beta = 1$, and $\theta$ is a fictitious time. The noises $\eta_\sigma$ and $\eta_\xi$ are complex Gaussian fields with zero mean. At each step, the fields are transformed to real space and combined into $\Phi(\mathbf{r}) = \xi(\mathbf{r}) - i\sigma(\mathbf{r})$, from which we estimate the normalized local weight $J(\mathbf{r}) = \exp[\Phi(\mathbf{r})] / \langle \exp[\Phi(\mathbf{r})] \rangle_{\mathbf{r}}$. For numerical implementation, we adopted the publicly available codes from Refs. 120 and 121. Readers are referred to Refs. 86, 118, 119 for detailed numerical algorithms and discussion of complex Langevin sampling in field-theoretic simulations.

### C. Structural Correlations

To assess how well structural correlations are encoded in the partition function, we estimated the radial distribution function (RDF), *g(R)*, from field sampling and compared it with a particle-level reference. For the reference RDF, we used the real-space potential $U(R)$ [Eq. (35)] and performed

Metropolis Monte Carlo with single-particle trial moves accepted according to $P_{\mathrm{acc}} = \min[1, \exp(-\beta\Delta E)]$. The RDF was computed by binning pairwise separations, and the result is shown in Fig. 1(c).

For the generalized mode theory, we sample the density correlation from the complex Langevin evolution and compute the structure factor $S(k) = \langle \rho(k)\rho(-k)\rangle / N$, from which the RDF is obtained via the 2D Hankel transform

$$g(R) = 1 + \frac{1}{\rho}\int_0^\infty \frac{k\,dk}{2\pi}[S(k) - 1]J_0(kR). \tag{41}$$

Figure 1(c) shows that the RDF estimated by the generalized mode theory exhibits good agreement with the particle-level reference. Minor discrepancies reflect numerical aspects of the field-based approach, such as spectral discretization and the requirement that the $k$-grid sufficiently resolves the separation between the positive-mode $\sigma(\mathbf{k})$ and the negative-mode field $\xi(\mathbf{k})$. A more systematic discussion of these numerical considerations, along with enhanced sampling strategies for generalized-mode decompositions, will be provided in a companion paper. Here, our primary aim is to demonstrate the feasibility of the generalized mode framework as a foundation for hierarchical bottom-up coarse-graining. The agreement in Fig. 1(c) confirms that the theory consistently captures both positive and negative Fourier modes, thereby validating its potential for constructing robust mesoscopic field representations.

## V. Hierarchical Coarse-Graining

### A. Atomistic → Molecular CG Level

Building on the theoretical developments in Secs. II and III, along with the supporting numerical demonstration in Sec. IV, we now introduce a hierarchical coarse-graining paradigm designed for addressing the complex nature of molecular interactions. We begin with the most detailed description of the system, i.e., the atomistic level. For molecules composed of $n$ particles, the atomistic configurations are denoted as $\mathbf{r}^n$, and the corresponding atomistic interaction, $u(\mathbf{r}^n)$, is typically expressed in molecular mechanics as:

$$u(\mathbf{r}^n) = \sum_{nb} u_{nb}(\mathbf{r}^n) + \sum_{b} u_b(\mathbf{r}^n) + \sum_{c} u_c(\mathbf{r}^n), \tag{42}$$

where *nb*, *b*, and *c* represent non-bonded, bonded, and charged interactions, respectively. As previously discussed, the divergent repulsive terms prevalent in $u_{nb}$ and $u_c$ present significant challenges for deriving a well-defined effective field-theoretic transformation.

To address this challenge, we construct a molecular CG model from Eq. (42) by representing the reference system with $N$ CG particles, where $N \ll n$, consistent with the earlier description in Sec. II. For simplicity, in this work, we represent each CG liquid molecule by its center-of-mass, denoted as $\mathbf{R}_I$, which is determined by the mapping operator $\mathbf{M}(\mathbf{r}^n)\colon \mathbf{r}^n \to \mathbf{R}^N$. Nevertheless, as we will discuss later in Subsection V B, the framework can readily accommodate multiple-site CG models without loss of generality. The effective CG interactions between these center-of-mass CG sites are assumed to be pairwise additive:

$$U(\mathbf{R}^N) = \frac{1}{2}\sum_{I \neq J} U(\mathbf{R}_I - \mathbf{R}_J),$$

$$(43)$$

where the pairwise CG potential $U(\mathbf{R}_I - \mathbf{R}_J)$ is only dependent on $R = |\mathbf{R}_I - \mathbf{R}_J|$. Comparing to Eq. (42), the relatively simple interaction form of Eq. (43), with fewer degrees of freedom, makes this approach advantageous for deriving the field-theoretic models. However, it is important to emphasize that the overall CG potential energy, $U(\mathbf{R}^N)$, should not be equated with the fine-grained (FG) internal energy $u(\mathbf{r}^n)$ from Eq. (1). Instead, $U(\mathbf{R}^N)$ approximates the FG free energy in the CG variables,[19-22] accounting for the "missing entropy" effect.[122-124] As demonstrated in particle-level coarse-graining, addressing the so-called *representability issue*[125-129] will become critical for accurately estimating field-based properties, which will be further explored in follow-up work.

The effective pairwise CG interaction $U(R)$ in Eq. (43) can be statistically inferred through various bottom-up coarse-graining methodologies. The main objective of bottom-up CG approaches is to determine effective equations of motion for the reduced representation that can capture important correlations at the CG level. Since forces are central to equilibrium equations of motion in both FG and CG variables, we adopt the MS-CG approach, which uses the force-matching methodology.[19-22, 31] In MS-CG, $U(\mathbf{R}^N)$ is determined variationally by minimizing the force residuals, $\chi^2[\mathbf{F}]$, between the FG and CG systems:

$$\chi^2[\mathbf{F}] = \frac{1}{3N}\left\langle \sum_{I=1}^{N} \left| \mathbf{F}_I(\mathbf{M}(\mathbf{r}^n)) - \mathbf{f}_I(\mathbf{r}^n) \right|^2 \right\rangle, \tag{44}$$

where $\mathbf{F}_I(\mathbf{M}(\mathbf{r}^n))$ is the target CG force that is designed for matching the projected atomistic forces on the CG site *I*, containing atoms in the set $\mathbb{I}_I$, $\mathbf{f}_I(\mathbf{r}^n)$ is defined as $\mathbf{f}_I(\mathbf{r}^n) \coloneqq \sum_{i \in \mathbb{I}_I} \mathbf{f}_i(\mathbf{r}^n)$. From the optimized pairwise force $\mathbf{F}_I(\mathbf{R}^N)$, the effective potential can be derived, as $\mathbf{F}_I(\mathbf{R}^N) = -\nabla_I U(\mathbf{R}^N)$. The pairwise forces obtained through this procedure satisfy the Yvon-Born-Green equation, thereby capturing up to three-body correlations.[130] The MS-CG approach is particularly well-suited for hierarchical coarse-graining, as the primary focus here is on structural (static) correlations. For a more comprehensive review of bottom-up CG methodologies, the reader is referred to recent articles.[17, 18]

**B. Molecular CG Level: Perturbation Theory in Reciprocal Space**

While molecular CG approaches eliminate extraneous degrees of freedom and capture essential structural features at reduced resolution, they do not remove the divergent short-range repulsions present at the atomistic level. This behavior is illustrated in Fig. 2 using both FG and CG interactions of a representative liquid, carbon tetrachloride ($CCl_4$). In the conventional atomistic force field [here, Optimized Potential for Liquid Simulations (OPLS)[131, 132]], the interactions are decomposed into non-bonded, bonded (bond and angle), and electrostatic components. Since bonded terms typically follow harmonic forms, we focus on the non-bonded and charged interactions, depicted in Figs. 2(a) and (b), respectively. These contributions are governed by Lennard-Jones and Coulombic potentials, both of which diverge at zero separation. Therefore, a rigorous Fourier transformation cannot be applied because $\int_{r=0}^{r=\infty} d\mathbf{r}\, u_\beta(\mathbf{r}) \exp(-i\mathbf{k}\cdot\mathbf{r})$ ($\beta$: *nb* and *c*) becomes ill-defined at $u_\beta(r=0)$ in the large wavenumber limit. This divergence is also manifested in the effective CG interactions derived from atomistic statistics: because configurations near the divergent region are never sampled, the corresponding regions of the CG potential remain undefined.

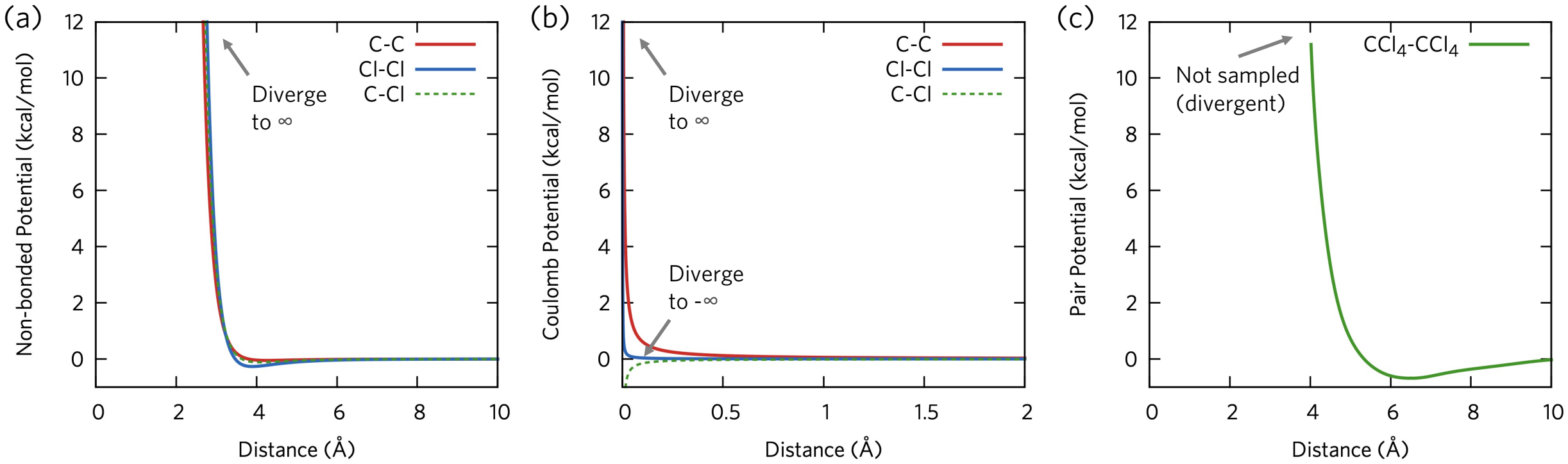

**Figure 2.** Divergent nature of conventional atomistic and CG force fields at the microscopic level, illustrated for $CCl_4$ at 300 K: (a) Non-bonded interactions between C and Cl pairs, represented by the Lennard-Jones potential. (b) Charged interactions between C and Cl pairs, described by the Coulomb potential. (c) Effective CG interaction. While coarse-graining removes high-frequency atomistic degrees of freedom, the resulting CG potential still contains short-range divergences that are not sampled during parameterization.

To address this divergence in CG interactions, one common strategy is to manually truncate the repulsive core to a finite value. However, without a systematic basis, such *ad hoc* truncation may neglect essential structural correlations and fail to generalize across diverse molecular systems. To resolve this, motivated by classical perturbation theory of liquids[133, 134] and its application to CG dynamics,[135-139] we introduce a perturbative expansion of molecular interactions in reciprocal space, as developed in Ref. 140. Building upon integral equation theory of molecular liquids, this formalism provides an approximate yet controlled representation of CG interactions in reciprocal space. At its core, the CG interaction potential is represented as:

$$U(R) = \beta^{-1}[h(R) - \ln g(R) - c(R) + B(R)], \tag{45}$$

where $h(R) = g(R) - 1$ is the total correlation function, $c(R)$ is the direct correlation function, and $B(R)$ is the bridge function. For simplicity, we adopt the hypernetted chain approximation, setting $B(R) = 0$. A more general treatment including $B(R)$ is discussed in Ref. 140 Expanding the logarithm term for low-density systems (a high-density generalization is provided in Ref. 140), we write

$$\ln g(R) = \sum_{n=1}^{\infty} (-1)^{n+1} \frac{[h(R)]^n}{n}. \tag{46}$$

Transforming into reciprocal space, $U(k)$ can be represented as the perturbative expansion:

$$U(k) = U^{(0)}(k) + \sum_{n=1}^{\infty} U^{(n)}(k), \tag{47}$$

where the zeroth-order term $U^{(0)}(k) = (\rho\beta)^{-1}[S(k)^{-1} - 1]$ corresponds to the random phase order approximation, and each perturbative correction is given by $U^{(n)}(k) = (-1)^n h^n(k)/(\beta n)$.

Each term in this expansion is finite and decays rapidly with increasing order $n$, thereby offering a systematic approach to regularize divergent CG interactions. In practice, the expansion is truncated at a finite order once the CG energetics have converged. To demonstrate this approach, we apply the perturbative expansion to carbon tetrachloride. At truncation order $n=10^5$, the resulting CG interaction accurately reproduces the total energetic profile and the center-of-mass RDF from the atomistic reference simulation (see Appendix C for truncation order analysis). As shown in Fig. 3(b), the perturbative expansion yields a finite reciprocal-space interaction $U(k)$, making it directly applicable within the generalized mode theory framework.

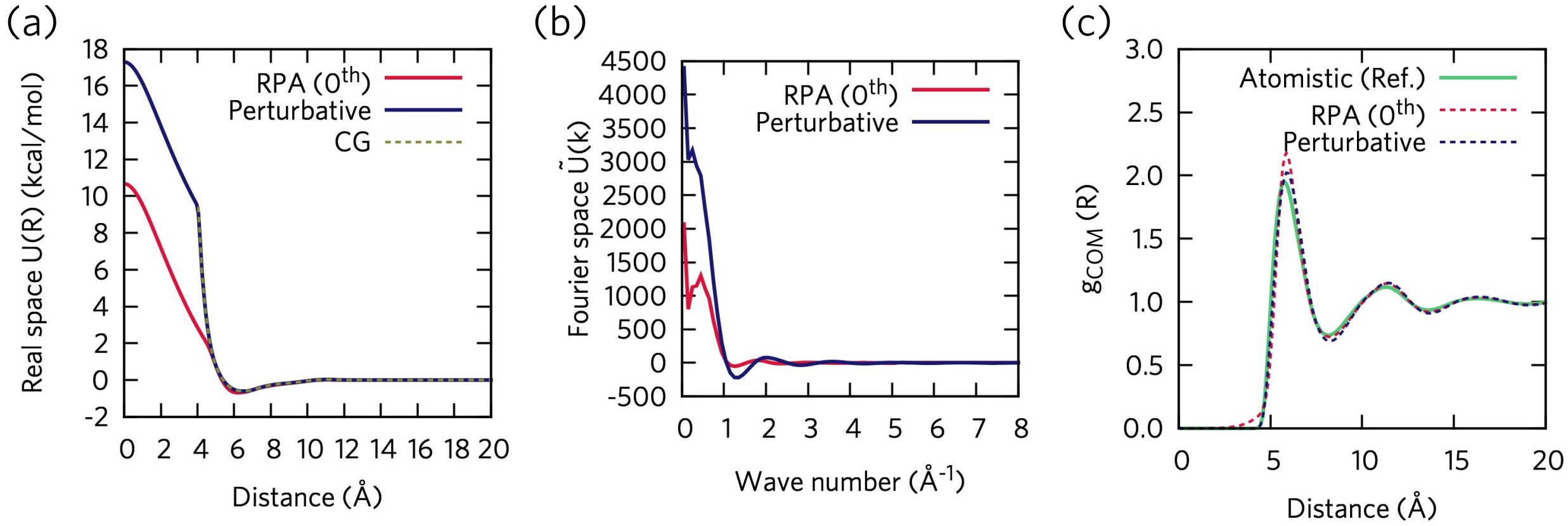

**Figure 3**. Perturbative expansion of the CG $CCl_4$ liquid in real space [panel (a)] and reciprocal space [panel (b)], showing the zeroth-order approximation (RPA; red solid), the perturbative interaction after $n=10^5$ orders (blue solid), and the real-space CG interaction from Fig. 2(c) (grey dashed). (c) The perturbative interaction in reciprocal space yields an improved center-of-mass RDF (blue dashed) relative to the RPA (red dashed), compared with the atomistic reference (green solid).

**C. Practical Approximation to Perturbative Interaction**

So far, we have shown that the generalized mode theory, combined with reciprocal perturbation theory, can effectively reformulate partition functions in reciprocal space. While, in principle, $\phi(\mathbf{k})$ can be uniquely decomposed into $\nu(\mathbf{k})$ and $\omega(\mathbf{k})$, the resulting profile for either $\nu(\mathbf{k})$ or $\omega(\mathbf{k})$ exhibits a discontinuous Fourier spectrum as it crosses zero. Since the HS transformation is applied separately to $\nu(\mathbf{k})$ and $\omega(\mathbf{k})$, this leads to two separate transformations being performed on spiky, discontinuous Fourier modes derived from molecular CG interactions. More importantly, the auxiliary fields $\sigma(\mathbf{k})$ and $\xi(\mathbf{k})$, which are sampled from these discontinuous Fourier profiles, may contribute to numerical issues, particularly when handling more complex systems such as multi-component systems with distinct interactions.

We further argue that this issue is exacerbated in molecular systems. In particular, (isotopic) pair interactions in liquids and other molecular systems typically exhibit short-range repulsions and comparatively weaker long-range interactions. As the pair distance increases, this interaction decays to zero, since simple liquids do not exhibit long-range ordering. Therefore, it is reasonable to deconvolute the pair CG interaction $U_2(R)$ into a convolution of two radial functions

$$U_2(R) = \sigma(R) * \widetilde{U}(R), \tag{48}$$

where $\sigma(R)$ is a sigmoid-like function that encodes the radially decaying feature of $U_2(R)$, and $\widetilde{U}(R)$ represents a deconvoluted $U_2(R)$ using the $\sigma(R)$ filter. For simplicity, $\sigma(R)$ can be approximated as a square step function that decays from 1 to 0 after some cutoff distance $R_{\text{cut}}$, which is related to the characteristic length scale of liquids:

$$\sigma(R) = \begin{cases} 1 \text{ (if } R < R_{\text{cut}}) \\ 0 \text{ (otherwise)} \end{cases}. \tag{49}$$

Since the Fourier transform of the square-like step function is the sinc function, $\text{sinc}(x) = \sin(x)/x$, this deconvolution directly reveals that the radially decaying nature of CG interactions in simple liquids (and in many molecular systems) inevitably leads to highly oscillatory Fourier modes in reciprocal space. Since the sinc function alternates in sign over a range of wave vectors, Eq. (49) implies that the Fourier transform of an "ordinary" molecular interaction will exhibit multiple sign changes across $k$-space. These frequent positive-to-negative mode transitions pose significant numerical challenges, particularly when discretizing wavevectors to separate these crossovers while still conforming to periodic boundary conditions.

To address the practical challenge of oscillatory Fourier-transformed CG interactions, we explore a practical approximation for the perturbative potential $U(k)$ of $CCl_4$, that aims to reduce numerical instability. The simplest approximation is naturally to retain only the low-$k$ (long-wavelength) modes, which do not exhibit oscillatory behavior and carry the dominant features of the interaction. In practice, we define a cutoff wavevector $k_0$ as the smallest nonzero value of $k$ at which $U(k)$ first crosses zero, i.e., $U(k_0) = 0$. The zeroth-orther approximation is then given by

$$U^{(0)}(k) = \begin{cases} U(k) \ (0 \le k \le k_0) \\ \quad 0 \ (k \ge k_0) \end{cases}, \tag{50}$$

which isolates the smoothly varying repulsive contribution by truncating higher-$k$ oscillations. To assess the fidelity of this zeroth-order approximation, we approximate the full Fourier-transformed interaction profile by focusing primarily on the short-range repulsive contributions in $k$ space, leading to $U^{\text{ref}}(k) \approx U^{(0)}(k)$ [Fig. 4(a)].

The fidelity of the zeroth-order approximation can be assessed by evaluating the inverse Fourier transformation of $U^{(0)}(k)$, i.e., $U^{(0)}(R) = (2\pi^2)^{-1} \int_0^\infty dk k^2 U^{(0)}(k) \sin(kR)/(kR)$, and then performing particle-level MD simulations using $U^{(0)}(R)$ [Fig. 4(b)]. While one could also directly perform field-theoretic simulations using $U^{(0)}(k)$, as outlined in Sec. IV, such simulations are inherently sensitive to numerical instabilities and sampling choices, particularly in the presence of highly oscillatory modes. These issues will be carefully examined in a companion paper. Instead, we use the MD simulation with $U^{(0)}(R)$ to provide an indirect benchmark for the maximum structural fidelity that could be expected if the reciprocal-space sampling were carried out exactly. In this way, the real-space simulation serves as a reference point for assessing the potential of the reciprocal-space approximation. We also note that $U(R)$ reconstructed numerically from $U(k)$ using a finite maximum wavevector must be interpreted with care, since a hard reciprocal-space truncation can generate oscillatory long-range artifacts (analogous to Gibbs ringing). In practice, the large-distance decay of $U(R)$ should be explicitly examined, and if oscillatory tails persist, additional numerical considerations, such as increased $k$-space resolution or smooth truncation schemes, may be employed.

From the inverse Fourier-transformed $U^{(0)}(R)$, we observe that omitting the higher-$k$ components slightly alters the pair interaction profile. While the low-$k$ component, corresponding to long-range distances, are reasonably preserved, the omission of high-$k$ modes affects the short-range behavior, particularly the attractive well between 4-8 Å. This is reflected in the particle-level RDF shown in Fig. 4(c). Although the general features, such as coordination shell distances and the position of the first peak, are well captured, deviations appear at both short distances and in overstructured regions at longer distances. These results indicate that truncating to only positive $k$-modes and applying a single HS transformation, as is commonly done in polymer field theory, may fail to capture important short-range features specific to molecular systems.

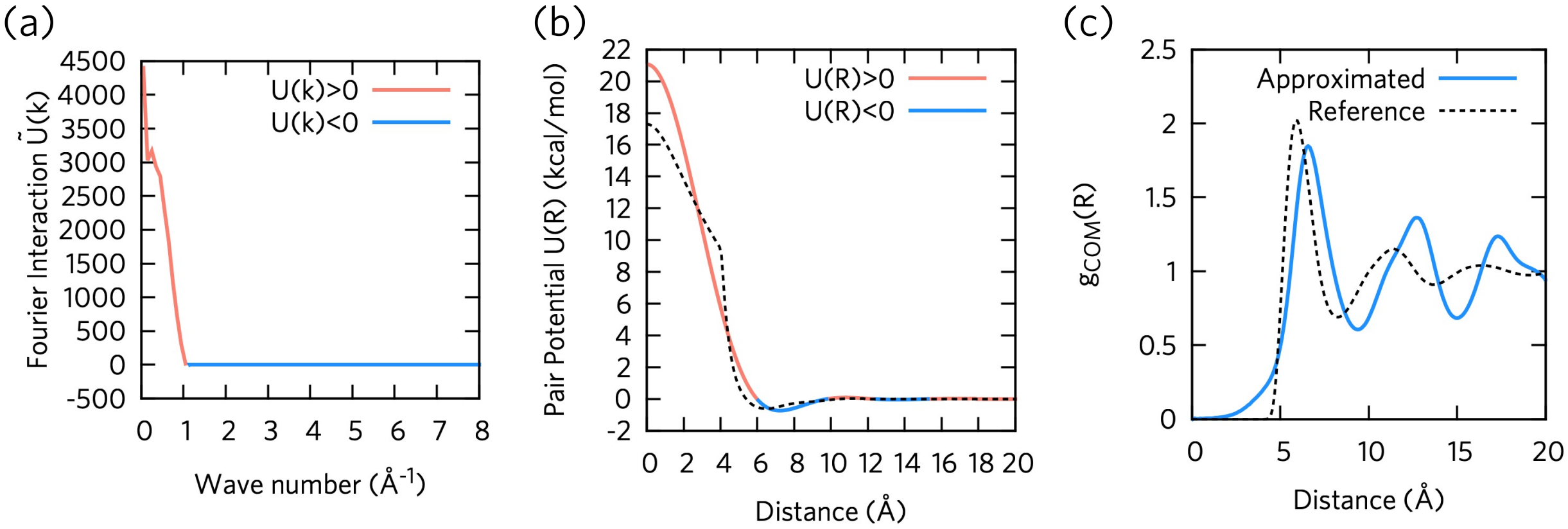

**Figure 4.** Zeroth-order approximation of the perturbative interaction for CG $CCl_4$ liquid in reciprocal space [panel (a)] and real space [panel (b)] with positive (red) and negative (blue) modes. The reference perturbative interaction from Fig. 3(a) is shown for comparison (black dashed). (c) Performance of the zeroth-order approximation assessed via particle MD simulation and the resulting RDF (blue solid), compared with the reference (black dashed).

To further improve the accuracy of this approximation, we next incorporate both positive and negative modes within the low-$k$ regime, constituting a first-order correction. In this case, we retain the positive-definite component $\nu(k)$ up to $k_0$, and the negative-definite component $\omega(k)$ up to a second cutoff $k_1$, where $k_0$ and $k_1$ are defined as the first and second nonzero values of $k$ for which $U(k_0) = U(k_1) = 0$. For $CCl_4$, this first-order approximation corresponds to keeping $\widehat{U}^{(1)}(k) \neq 0$ for $k = k_1 = 1.75$ Å$^{-1}$, as shown in Fig. 5(a), and setting $U(k) = 0$ for $k > k_1$. This first-order approximation introduces a second auxiliary field but avoids numerical instability by excluding contributions from $k > 1.75$ Å$^{-1}$. The resulting real-space interaction profile, Fig. 5(b), shows that even a minimal inclusion of negative modes enhances the attractive well compared to the zeroth-order case in Fig. 4(a). Although the interaction deviates at extremely short distances, these regions are well beyond thermal energy ($\gg k_B T$) and are rarely sampled in MD simulations.

When we assess the first-order approximation via particle-level MD, we find improved structural correlations at both short ranges (< 5 Å) and long ranges ( > 10 Å). While some deviations remain in RDF intensities, the coordination shell positions are accurately reproduced. Together, Figs. 4 and 5 demonstrate that retaining only the short-wavelength (low-$k$) modes for both positive and negative components can offer a practical approximation for evaluating structural properties in auxiliary field simulations, while mitigating the numerical instabilities associated with highly oscillatory reciprocal-space interactions.

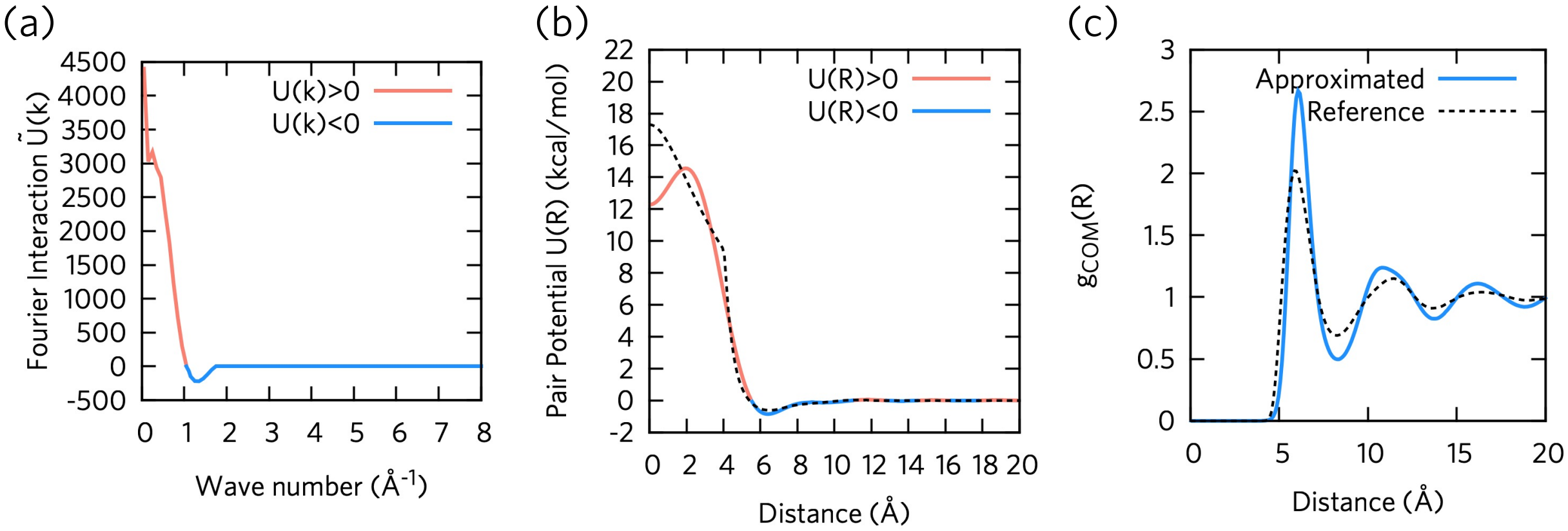

**Figure 5.** First-order approximation of the perturbative interaction for CG $CCl_4$ liquid in reciprocal space [panel (a)] and real space [panel (b)] with positive (red) and negative (blue) modes. The reference perturbative interaction from Fig. 3(a) is shown for comparison (black dashed). (c) Performance of the first-order approximation assessed via particle MD simulation and the resulting RDF (blue solid), compared with the reference (black dashed).

## VI. Conclusions

To conclude, we propose a new numerical framework for constructing bottom-up auxiliary field representations of arbitrary (and potentially complex) molecular systems. This combined CG framework proceeds in two key stages. First, we apply a bottom-up molecular CG approach to reduce the complexity of the atomistic system to a center-of-mass representation (Fig. 2). Although this CG step removes high-frequency atomistic degrees of freedom, the resulting interactions still retain strong short-range divergences. To regularize these divergence, we introduce a perturbative expansion in reciprocal space, which allows us to systematically control the perturbation order and approximate the ill-defined reference potential with a convergent, well-behaved one (Fig. 3). In the second stage, we perform two independent HS transformations on the regularized potential, separating the positive and negative Fourier modes and introducing two auxiliary fields. Motivated by advances in polymer field-theoretic simulations, including work by Fredrickson, Shell, and others, we generalize the formal derivation of Baeurle and Parinello, which was previously restricted to positive-definite modes, to arbitrary mode decompositions applicable to both canonical and grand canonical ensembles at the CG level. In doing so, we also correct inconsistencies in earlier formulae.

We demonstrate that the generalized mode theory with existing field-theoretic sampling techniques (e.g., complex Langevin sampling) can reproduce key structural correlations, consistent with conventional mean-field SCFT approaches (Fig. 1). However, in contrast to the simpler positive-only case, generalized modes introduce highly oscillatory features in reciprocal space, exacerbating both the numerical instability and the intrinsic phase problem associated with the HS transformation. We show that restricting the sampling to positive and negative modes in short-*k* regimes mitigates these challenges while retaining essential structural features (Fig. 5). This numerical protocol, summarized in Fig. 6 below, extends the reach of the bottom-up coarse-graining to molecular systems with nontrivial pair (and bonded) interactions, offering a systematic route to field-theoretic reresentations. While a single, fully integrated example demonstrating the complete pipeline from molecular liquid to field-theoretic sampling is not provided here, owing to the remaining numerical challenges in generalized mode sampling, this work lays the formal and numerical groundwork necessary for such future efforts. Building on the findings of this work, a follow-up study will address these issues in detail.

By unifying formal derivations, numerical strategies, and insights from the polymer field-theory community building on ongoing efforts,[97-100] we anticipate that this hierarchical coarse-graining framework provides a scalable and generalizable foundation for field-theoretic modeling of molecular systems, enabling predictive and multiscale simulations of complex condensed-phase behavior.

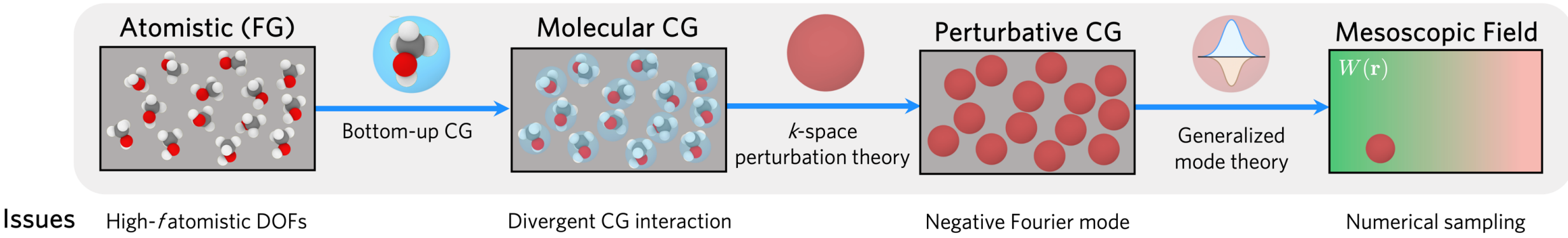

**Figure 6.** Schematic overview of the bottom-up hierarchical coarse-graining framework from atomistic to molecular CG, to perturbative CG, and finally to a mesoscopic auxiliary field representation.

## ACKNOWLEDGMENTS

This material is based upon work was supported by the National Science Foundation (NSF Grant CHE-2102677). Simulations were performed using computing resources provided by the University of Chicago Research Computing Center (RCC). J.J. gratefully acknowledges Dr. Ankit Mahajan and Professor David R. Reichman for insightful discussion and feedback. J.J. also thanks the Harper Dissertation Fellowship from the University of Chicago during his graduate studies and an Arnold O. Beckman Postdoctoral Fellowship (http://dx.doi.org/10.13039/100000997) during his postdoctoral appointment.

## DATA AVAILABILITY

The data that support the findings of this work are available from the corresponding author upon request.

## APPENDIX

### A. Treatment of Bonded Interactions in Multiple-Site CG Models

In this Appendix, we demonstrate how intramolecular (bonded) interactions remain decoupled from the HS transformation. To effectively illustrate our perspective, we showcase bonded interactions for two-site models and three-site CG models with bond and angle potentials. The reasoning provided can be faithfully extended to longer chains or more complex intramolecular topologies.

#### 1. Bonded interactions

Consider a two-site CG model where each molecule (total *N*) comprises two sites labeled $\mathbf{R}_{2I}$ and $\mathbf{R}_{2I+1}$ for the *I*-th molecule. We assume that the bonded potential between these CG sites follows a simple harmonic form:

$$U_{\text{bond}} = \sum_{I=1}^{N} \frac{k_b}{2} |\mathbf{R}_{2I} - \mathbf{R}_{2I+1}|^2, \tag{A1}$$

where $k_b$ is the bond constant. To see how bonded terms can be analytically integrate out, consider a single bond

$$u_{\text{bond}} = \frac{k_b}{2} |\mathbf{R}_2 - \mathbf{R}_1|^2. \tag{A2}$$

The bond configuration integral over $\mathbf{R}_1$ and $\mathbf{R}_2$ is then

$$\mathcal{Z}_{\text{bond}}^{1-2} = \int d\mathbf{R}_1 d\mathbf{R}_2 \exp\left(-\frac{\beta k_b}{2} |\mathbf{R}_2 - \mathbf{R}_1|^2\right), \tag{A3}$$

where $\beta = (k_B T)^{-1}$. By performing a change of variables from the canonical transformation $\bar{\mathbf{R}} := (\mathbf{R}_2 + \mathbf{R}_1)/2$ and $u := \mathbf{R}_2 - \mathbf{R}_1$, the integral over $\bar{\mathbf{R}}$ gives a factor of the total volume $V$, whereas the integral over $u$ yields $[2\pi/(\beta k_b)]^{3/2}$. Combined these, we arrive at $\mathcal{Z}_{\text{bond}}^{1-2} = V[2\pi/(\beta k_b)]^{3/2}$, and the total bond configuration integral is

$$\mathcal{Z}_{\text{bond}} = V^N \left(\frac{2\pi}{\beta k_b}\right)^{\frac{3N}{2}}, \tag{A4}$$

which contributes as a Gaussian factor to the overall partition function, independent of the HS decoupling step.

**2. Angle interaction**

For three-site CG molecules, the same logic extends to both bond lengths and angles. Suppose that each of the *N* molecules has sites $\mathbf{R}_{3I}$, $\mathbf{R}_{3I+1}$, and $\mathbf{R}_{3I+2}$. The intramolecular interactions for this three-site system can be expressed as:

$$U_{\text{intra}} = \sum_{I=1}^{N} \left[\frac{k_b}{2} |\mathbf{R}_{3I+1} - \mathbf{R}_{3I}|^2 + \frac{k_b}{2} |\mathbf{R}_{3I+2} - \mathbf{R}_{3I+1}|^2 + \frac{k_\theta}{2} |\theta_I - \theta_0|^2\right], \tag{A5}$$

where $\theta_I$ is the angle formed by the two bonds in the $i$-th molecule with the energy minimum angle $\theta_0$, and $k_\theta$ is an angle constant. Although the integral in Eq. (A5) is more complex than Eq. (A1) due to angular term, the analytical integral follows a similar structure. By transforming to appropriate internal coordinates [e.g., bond-angle-torsion (BAT) coordinates[141-144]] to handle bond lengths and angles, Eq. (A5) remains a Gaussian-like integral in bond distances, multiplied by an angular term. Importantly, none of the bond or angle terms in Eq. (A5) follow the density-density interaction form $\rho(\mathbf{k})\rho(\mathbf{k}')$ required by the HS transformation. Therefore, the HS transformation applied to non-bonded interactions for three-site molecules does not affect these intramolecular constraints.

In summary, for an arbitrary number of internal constraints (bonds, angles, dihedrals, etc.), these bonded interactions remain in the direct coordinate integral and do not require HS transformation. Consequently, for complex CG models involving many CG sites with non-bonded and bonded interactions, the resulting partition function naturally separates into an analytical integral over intramolecular coordinates and the HS-transformed portion of the free energy, capturing intermolecular interactions in a field representation. This separation ensures that multi-site CG liquids and complex molecular topologies can be incorporated into our framework without any loss of generality.

**B. Computational Details**

The atomistic and CG models for liquid carbon tetrachloride were constructed based on our previous work.[145] The initial structure, comprising 1000 $CCl_4$ molecules, was arranged in a cubic box with a length of 50 Å. The Packmol program package[146] was used to randomize the initial configuration, followed by energy minimization to remove artificial stresses introduced during system preparation. To equilibrate the system, constant *NPT* simulations were performed at 300 K and 1 atm for 1 ns. From the equilibriated configuration, the box size was fixed at 57.23 Å, and constant *NVT* simulations were conducted at 300 K for 5 ns using the Nosé-Hoover thermostat.[147, 148] All MD simulations were carried out with the Large-scale Atomic/Molecular Massively Parallel Simulator (LAMMPS) program package.[149-151]

From the sampled *NVT* trajectory, we coarse-grained each $CCl_4$ molecule to its center-of-mass representation. At this reduced resolution, the effective CG interaction was determined using the force-matching technique, implemented in the OpenMSCG program suite.[152] Sixth-order B-spline functions with a resolution of 0.2 Å were employed to obtain the final interaction profiles. The perturbative framework and numerical implementation followed the procedures described in Ref. 140.

**C. Systematic Truncation of Perturbative Orders**

The perturbative expansion for molecular systems introduced in Sec. V B leads to an effective interaction kernel in reciprocal space that can be truncated at a finite order $m$,

$$U_m(k) = U^{(0)}(k) + \sum_{n=1}^{m} U^{(n)}(k). \tag{C1}$$

In practice, the optimal truncation order $m$ can be systematically identified by estimating the approximate total potential energy in reciprocal space:

$$U_{\text{approx}}(\mathbf{R}^N) = \frac{1}{2}\sum_{k} \rho(k) U_m(k) \rho(k) - U_{\text{self}}^{m}, \tag{C2}$$

where $\rho(k)$ is sampled numerically from the mapped CG trajectory, and $U_{\text{self}}^{m}$ is well-defined since $U_m(k)$ remains finite. The approximate energy $U_{\text{approx}}(\mathbf{R}^N)$ can then be compared with the reference CG potential energy using $U(R)$ via $U_{\text{ref}}(\mathbf{R}^N) = 0.5 \sum_{I \neq J} U(R_{IJ})$ to determine an optimal truncation order.

Figure 7 illustrates this perturbation-order analysis for $CCl_4$ discussed in Sec. V B. We find that truncation at approximately $n \approx 100$ already provides an accurate representation of the reference system. Note that in the present work, we nevertheless include up to $n = 10^5$ to ensure completeness and to fully capture the interaction profile.

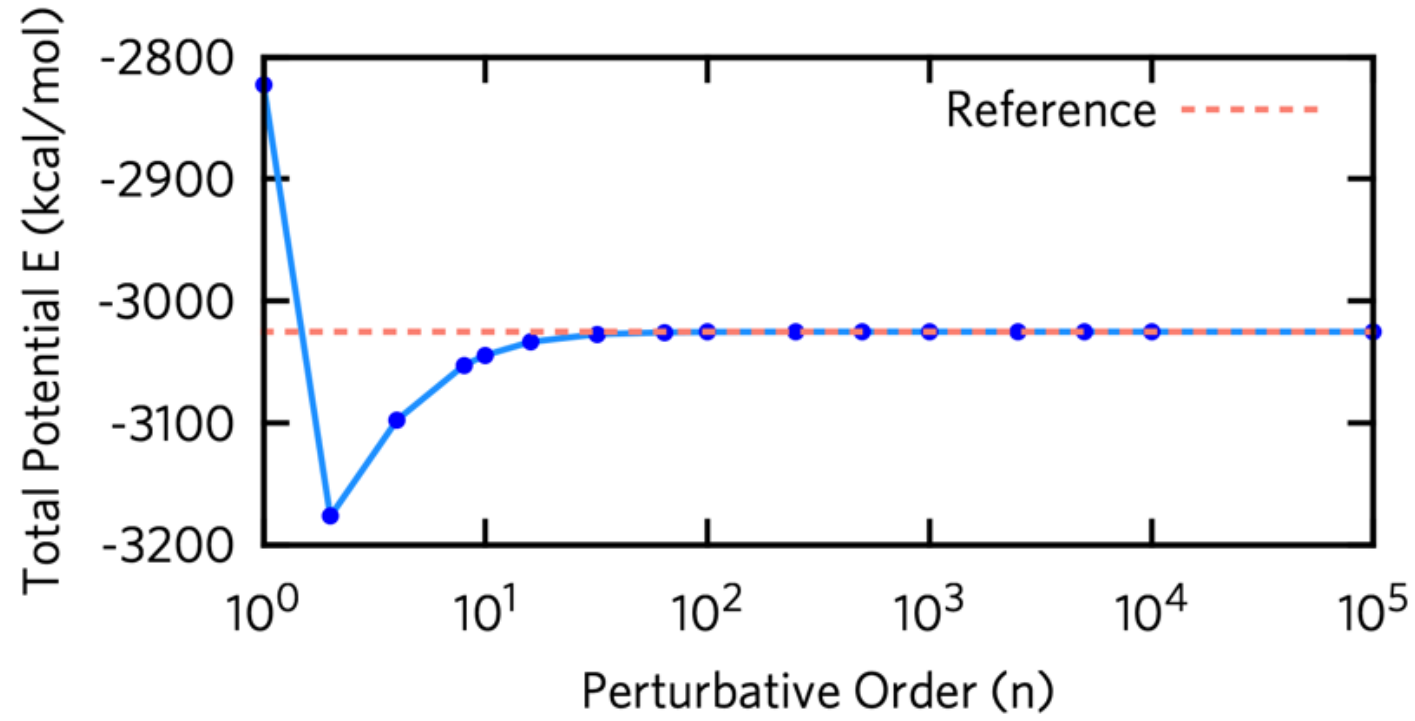

**Figure 7.** Determination of the truncation order for liquid $CCl_4$. The perturbative expansion [Eq. (C1)] approaches the reference potential energy at higher orders, with convergence observed for $n \geq 100$.